\documentclass[11pt,a4paper]{article}

\usepackage[utf8]{inputenc}
\usepackage[english]{babel}

\usepackage{geometry}
\usepackage{graphicx}
\usepackage{amssymb,amsmath,amsthm}
\usepackage{stmaryrd}
\usepackage{delarray}
\usepackage{enumitem}
\usepackage{tikz}
\usepackage{hyperref}
\graphicspath{{figures/}}
\usepackage{authblk}
\usepackage{wrapfig}
\usepackage{multicol}
\usepackage{float}
\usepackage{appendix}
\usepackage{pdfpages}

\newtheorem{theorem}{\textbf{Theorem}}
\newtheorem*{theorem*}{\textbf{Theorem}}

\newtheorem{proposition}{\textbf{Proposition}}

\newtheorem{remark}{\textbf{Remark}}

\newtheorem{remark*}{Remark}
\newtheorem{example}{\textbf{Example}}

\renewcommand{\int}[1]{[#1]}
\newcommand{\intz}[1]{\llbracket#1\rrbracket}
\newcommand{\N}{\mathbb{N}}

\renewcommand{\O}{\mathcal{O}}

\newcommand{\Poly}{{\mathsf{P}}}
\newcommand{\NP}{{\mathsf{NP}}}
\newcommand{\coNP}{{\mathsf{coNP}}}
\newcommand{\PSP}{{\mathsf{PSPACE}}}

\newcommand{\dyna}[1]{\mathcal{G}_{#1}}

\newcommand{\decisionpbw}[4]{
	\fbox{\parbox{{#4}\textwidth}{{\bf {#1}}\\{\it Input:} {#2}\\{\it Question:} {#3}}}
}

\newcommand{\psiD}{\textbf{$\psi$-dynamics}}
\newcommand{\psiDq}{\textbf{$\psi$-$q$-AN-dynamics}}
\newcommand{\psiNDq}{\textbf{$\psi$-$q$-NAN-dynamics}}
\newcommand{\SAT}{\textbf{SAT}}
\newcommand{\CVP}{\textbf{CVP}}

\title{Circuit metaconstruction in logspace for Rice-like complexity lower bounds in ANs and SGRs}
\author[$\dag$]{Aliénor Goubault--Larrecq}
\author[$\dag$]{Kévin Perrot}
\affil[$\dag$]{Aix Marseille Univ, CNRS, LIS, Marseille, France}
\date{}
\begin{document} 
	\maketitle
	
        \begin{abstract}
          A new proof technique combining finite model theory and dynamical systems
	  has recently been introduced to obtain general complexity lower bounds
          on \emph{any} question one may formulate on the dynamics (seen as a graph)
          of a given automata network (AN).
          ANs are abstract finite dynamical systems of interacting entities
          whose evolution rules are encoded as circuits,
          hence the study also applies to succinct graph representations (SGRs).
	  In this article, we detail the construction of circuits
          to obtain general complexity lower bounds (metareduction)
          and show that the reduction is feasible in logarithmic space.

        \end{abstract}

	\section{Introduction}
	
	Automata networks (AN) constitute a general framework for modeling interacting entities. 
	They are widely used to model complex biological systems by 
	representing interactions among components such as genes, proteins, and cells. They 
	provide a powerful framework to study key dynamical properties happening in
	biological processes~\cite{k69,t95,m98,e04,ks08}.
	
	Each automaton has a state from a finite alphabet, and 
	updates it based on the states of others.
    Aggregating the states of all automata results in a configuration of the network.
    When the dynamics is deterministic, a
	transition function on the space of configurations can always be defined by local update 
	functions that are applied synchronously in discrete time steps. 
	Non-deterministic dynamics also exist and we usually have a transition relation between
	configurations but it is not always possible to compute
	the following state of each automaton independently.
	We consider automata networks of finite size,
	hence with a finite number of configurations.
    The dynamics of an automata network is
	represented as the graph of its transition function or relation
    (the vertices are the configurations).
	While various update modes exist, this article considers the fully synchronous 
	(parallel) update mode.
	This model is general, as any directed graph is the 
	dynamics of some automata network 
	(with out-degree 1 when it is deterministic).
	
	An automata network can be deterministic or non-deterministic,
        uniform (\emph{e.g.}~Boolean) or non-uniform,
        and it may follow various update policies.
	This allows automata networks to capture diverse dynamical behaviors 
	across different contexts~\cite{ks08,psa16,t73,ps22,lr23}.
	Beyond their biological applications, automata networks are also studied as 
	computational models in theoretical computer science, their mathematical properties and 
	algorithmic complexity being in the center of an active research 
	field~\cite{a85,abgs23,bgmps21,gmrwt21,sga22}.
	
	It is possible to express properties on the dynamics in the form of logical formulas. 
	Given a formula $\psi$ and an automata network $F$, 
	knowing whether or not the dynamics of $F$ (succinctly encoded in a circuit) 
	verifies the property
	$\psi$ is a theoretical problem
	meaningful for biological applications \cite{a85,bgmps21,ggpt21,ggpt23}.
	The question arises of how much time and space is necessary to compute its answer.
        For example, it has been proven that
	knowing if a dynamics contains a fixed point or not is 
	$\NP$-complete~\cite{a85}.
        An extension to the existence of a limit
        cycle of size $k$ has been shown to be $\NP$-complete as well~\cite{bgmps21}. 
        Other problems, still expressible as monadic second order formulas,
	have been studied. Questioning whether the update function of an automata network is 
	a bijection, the identity or a constant are $\coNP$-complete~\cite{p22}; and whether this
	function is nilpotent or not is $\PSP$-complete~\cite{ggpt21}.
	
        Inspired by Rice's theorem, which highlights the undecidability of \emph{any} non-trivial semantic
	property of programs, recent works have explored complexity-theoretic analogs for automata 
	networks. Since these systems are finite, the focus shifts from computability to 
	complexity. The aim of Rice-like metatheorems in this context is to show a clear
	dichotomy among the complexity of deciding properties.  
	The first metatheorem, generalizing previous results, proved that in the 
	deterministic case, AN properties expressible in first order logic are either trivial, $\NP$-hard 
	or $\coNP$-hard to verify~\cite{ggpt21}. A second metatheorem proved a similar result
	in the non-deterministic case and with properties expressible in monadic second order
	logic, hence it includes more formulas~\cite{ggpt23}. 
	A third metatheorem combined both techniques,
	for both deterministic and non-deterministic dynamics, at the level of monadic second order logic.
        A major improvement was to also fulfill a uniformity restriction,
	where all the automata have the same alphabet.
	The study focused on building models (that verify the property), and counter-models
	(that do not verify the property), using similar tools~\cite{glp24}.

	Based on these results, in this work we
        will detail the construction of circuits involved in these metareductions
        (reductions to obtain general complexity lower bounds for the metatheorems),
	while proving that it can be done in logspace (instead of the standard polytime considered so far).
	For hardness proofs, we reduce from the canonical problem $\SAT$.
        We provide an explicit view of the circuits produced by the metareductions,
        including the new developments of~\cite{glp24}.
        It will also serve as a cornerstone to transfer the metatheorems to other models of computation.
        The main theorem that we will prove in this article is (the notions are formally defined below):
	\begin{theorem}
		\label{theorem:main}
		Let $q\geq 2$.
                For any $q$-non-trivial (resp.~$q$-arborescent) MSO formula,
                the problem {\bf $\psi$-$q$-AN-dynamics} (resp.~{\bf $\psi$-$q$-NAN-dynamics})
                is $\NP$-hard or $\coNP$-hard for logspace reductions.
	\end{theorem}

        \paragraph{Outline} In Section~\ref{s:def}, we define the main concepts:
        automata networks, monadic second order logic, 
        and the family of decision problems proven to be hard. In Section~\ref{s:metareduction},
	we recall the models and counter-models constructed in~\cite{glp24},
	and provide all the necessary background and notations
        (the correctness of the metareduction is exposed in~\cite{glp24}).
	In Section~\ref{s:logspace},
	we detail how to compute the metareduction, \emph~{i.e.}
	how to construct the circuit of an automata network
        from a $\SAT$ formula
        (for both deterministic ANs and non-deterministic NANs). 
        A fine study allows to conclude that logarithmic space is enough (Theorem~\ref{theorem:main}).
        We conclude and present perspectives in Section~\ref{s:conclusion}
	
	\section{Definitions}
	 \label{s:def}
	
	For $n,q \in \N_+$, we denote 
	$\int{n} = \{1,\dots,n\}$ and $\intz{q}=\{0,\dots,q-1\}$. For a graph $G$, we denote
	$V(G)$ is its set of vertices, $E(G)$ its set of arcs, and $|G| = |V(G)|$. 
	
	\paragraph{Circuits}	
	A Boolean \emph{circuit} is an acyclic graph with a type associated to every vertex:
        either input, output, $\wedge$, $\vee$ or $\neg$ (or usual syntactic sugars).
	The size of a circuit is its number of vertices also called \emph{gates}.
	Its arcs are also called \emph{wires}.
	Inputs are source vertices, and outputs are sink vertices.
	A circuit computes a function from input words (one bit per input vertex)
	to output words (one bit per output vertex) with the standard semantics.
        It can simply be encoded by its adjacency matrix,
        together with a map associating a type to each gate vertex.

	\paragraph{Automata networks}
	A \emph{deterministic automata network} (AN) $F$ of size $n$ contains $n$ automata 
	where the $i^{th}$ automaton (where $i \in [n]$) has a state from its finite alphabet
        $A_i = \intz{q_i}$ for some $q_i\in\N_+$. The AN is a function 
	$F:X \to X$ where $X = \prod_{i \in [n]}A_i$ is the set of configurations of the AN.
        A \emph{non-deterministic automata network} (NAN) $F$ of size $n$
	is a relation $F \subseteq X \times X$. 
	An AN $F$ can be split into local functions $F_i: X \to A_i$ for each automaton $i \in \int{n}$,
        such that for any configuration 
	$x \in X, F(x) = (F_1(x), F_2(x), \dots, F_n(x))$.
        In the non-deterministic case (NAN), 
	we do not suppose the existence of local relations. 

	An automata network (AN or NAN) $F$ can be represented as a digraph called its
	\emph{dynamics} and denoted $\dyna{F}$, such that $V(\dyna{F}) = X$ and 
	$E(\dyna{F}) = \{(x, F(x)) | x \in X\}$ in the deterministic case, or $E(\dyna{F}) = F$ in the 
	non-deterministic case.
	When there is $q \in \N_+$ such that $q_i=q$ for all $i \in \int{n}$, 
	meaning that all automata have the same alphabet, 
	we say that $F$ is \emph{$q$-uniform}. In this case we have $X = \intz{q}^n$. 
	When $q = 2$, $X = \{0, 1\}^n$ and we say that $F$ is a \emph{Boolean} 
	automata network.

	The encoding of an automata network plays an important role here, given our focus
	on computational complexity. 
	An AN $F$ takes as input a configuration and outputs another configuration, hence it
	is encoded as a Boolean circuit with $\lceil \log_2 |X| \rceil$ input bits and 
	$\lceil \log_2 |X| \rceil$ output bits. A NAN $F$ takes as inputs two configurations
	$c$ and $c'$ and may or may not have a transition from $c$ to $c'$, hence
	it is encoded as a Boolean circuit with
	$2 \lceil \log_2 |X| \rceil$ input bits and 1 output bit. 
	
\begin{figure}[t]
		\centering
		\includegraphics{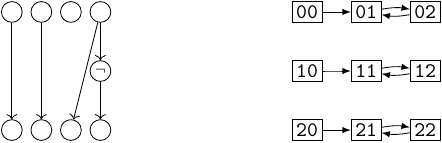}
		\caption{
            Example deterministic AN of size $n=2$
            on alphabet $\intz{q}=\{0,1,2\}$ with $q=3$.
            Circuit of the function $F:\intz{q}^n \to \intz{q}^n$
            (left)
            and transition digraph dynamics $\dyna{F}$ on configuration space $\intz{q}^n$ (right).
            A configuration is encoded on four bits, where inputs $0000$ to $0001$ correspond
            respectively to configurations $00$ to $22$.If we write a configuration of this AN
            $x_1 x_2 \in \intz{q}^2$, a way to define the local functions is
            $F_1(x_1 x_2) = x_1$, $F_2(x_1 x_2)= (x_2 \mod 2) +1$.
		}
		\label{fig:AN}
	\end{figure}

	\paragraph{Succinct graph representations} 
        The circuit encoding of a graph, namely $\dyna{F}$, is also called a
        \emph{succinct graph representation} (SGR)~\cite{gw83}.
        This forgets about the internal network structure made of interacting entities
        called automata.
        If $G$ is a succinct graph representation of the dynamics of $F$, we write $G(c)$ in the 
	deterministic case for the configuration such that $F(c) = G(c)$, and
	$G(c, c')$ in the non-deterministic case for the Boolean saying whether 
	$(c, c') \in F$ or not.
        See an example on Figure~\ref{fig:AN}.

	\paragraph{Tree-decompositions and treewidth}
	Tree-decompositions are widely used for the construction of graphs in~\cite{glp24}.
        Given $G = (V, E)$, 
	its tree-decomposition $T= (V_T, E_T)$ is a 
	tree such that every node $t \in V_T$ is associated to a bag 
	$\mathcal{B}(t) \subseteq V$ of vertices of $G$ verifying the following three properties:
        \emph{(1)} For every $v \in V$, there exists a node $t \in V_T$ such that $v \in \mathcal{B}(t)$,
        \emph{(2)} For every $(u, v) \in E$, there exists a node $t \in V_T$ such that $u \in \mathcal{B}(t)$
	    and $v \in \mathcal{B}(t)$,
        \emph{(3)} For any $v \in V$, if $T_v$ is the sub-tree of $T$ induced by the nodes whose
	    bag contains $v$ then $T_v$ is connected.
	The \emph{width} of a tree-decomposition $T$ is $w(T) = \max\{|\mathcal{B}(t)| \mid t \in V_T\} -1$. 
	A graph may admit several tree-decompositions of various widths but the most 
	interesting decompositions are the one of minimal width.
        The \emph{treewidth} of $G$ is defined as
	$tw(G) = \min \{w(T) \mid T \textnormal{ a tree-decomposition of } G \}$ \cite{df13}.

	\paragraph{Monadic second order graph logic}
        Our signature is $\{=, \to \}$
        where $\to$ is the binary relation (arcs) of a graph,
        \emph{i.e.}~for $x, y \in V(G)$ the atom $x \to y $ is true if and
	only if $(x, y) \in E(G)$. 
        Monadic second order (MSO) formulas are defined inductively by:
        $\psi = x \to y \mid x = y \mid x \in X \mid \exists x, \psi \mid \exists X, \psi \mid \psi \vee \psi \mid \neg \psi$,
        with all the usual syntactic sugars ($\forall, \wedge, \implies$),
        and where the quantification ($\exists$ or $\forall$) are 
        on vertices or sets of vertices (usually we denote a vertex $x$ and a set of vertices $X$).
	Given an MSO formula $\psi$, we say that a graph $G$ is a \emph{model} of $\psi$ and write 
	$G \vDash \psi$ when the graphical property described by $\psi$ is true in $G$.
        If it is not, we say that $G$ is a \emph{counter-model} and write $G \nvDash \psi$ 
        \cite{l04}.
        
        \begin{example}
        	For the AN $F$ presented in Figure~\ref{fig:AN} and the two formulas
        	$\psi = \exists x, x \to x$ (existence of a fixed point) and 
        	$\varphi = \exists x, \exists y, x \to y \wedge y \to x$ (existence of a cycle of length $2$), we have 
        	$\dyna{F} \nvDash \psi$ and $\dyna{F} \vDash \varphi$.
        \end{example}
	
	With our focus on computational complexity, we need to distinguish families of MSO
	formulas and especially define what we mean by triviality. 
	A formula $\psi$ is \emph{non-trivial} when it has infinitely many models and infinitely 
	many counter-models, among graphs of out-degree 1. A formula $\psi$ is 
	\emph{arborescent} when there exists $k$ such that $\psi$ has infinitely many models of 
	treewidth at most $k$ and infinitely many counter-models of treewidth at most $k$.
	In the restriction to graphs of out-degree 1, $\psi$ is either non-trivial
        or it has a finite number of models or a finite number of counter-models (then it is called 
        \emph{trivial}). However, without any graph restriction in the non-deterministic case, 
        when $\psi$ is not arborescent, it does not mean that $\psi$ is trivial.
	We also restrict formulas to $q$-uniform automata networks,
        which is equivalent to considering only graphs $\dyna{F}$ of size $q^N$ for some
	$N \in \N_+$.
        We obtain the notions of \emph{$q$-non-trivial} 
	and \emph{$q$-arborescent} formulas.

        \paragraph{Problems} 
        We will consider only the decision problems restricted to $q$-uniform networks,
        because the method identically applies to the weaker results without this restriction.
        We have two problems: one for the deterministic case (AN)
        and one for the non-deterministic case (NAN).
	
	\smallskip
	\noindent
	\decisionpbw{$\psi$-$q$-AN-dynamics}
	{circuit of a $q$-AN $f$ of size $N$.}
	{does $\dyna{f} \vDash \psi$?}
	{.45}
	\decisionpbw{$\psi$-$q$-NAN-dynamics}
	{circuit of a $q$-NAN $f$ of size $N$.}
	{does $\dyna{f} \vDash \psi$?}
	{.47}
	\smallskip

	
	\paragraph{Boundaried graphs}
	A \emph{boundaried graph} $G$ 
        has two tuples of
	ports of the equal size: the primary ports $P_1(G) \subseteq V(G)$ and the secondary ports $P_2(G) \subseteq V(G)$.
	In general, we do not suppose that $P_1(G) \cap P_2(G) = \emptyset$, but we always have
	$|P_1(G)| = |P_2(G)|$. In the case where $k = |P_1(G)|$ we call $G$ a 
	\emph{$k$-boundaried graph}, or \emph{$k$-graph} for short.
	Given two $k$-graphs $G$ and $G'$, we call 
	\emph{gluing} the operation of taking their disjoint union, and merging
        the secondary ports of the first graph to the primary ports of the second graph.
	In is denoted $G \oplus G'$, and is formally the $k$-graph defined as: 
	\begin{align*}
		V(G \oplus G') &= V(G) \cup (V(G') \backslash P_1(G')),\\
		E(G \oplus G') &= E(G) \cup E(G'[P_2(G)\leftarrow P_1(G')]),\\
		P_1(G \oplus G') &= P_1(G),\\
		P_2(G \oplus G') &= P_2(G'),
	\end{align*}
	where if $P_2(G) = (p_1, \dots, p_k)$ and $P_1(G') = (p'_1, \dots, p'_k)$ then 
	$G'[P_2(G)\leftarrow P_1(G')]$ denotes the graph $G'$ where each vertex $p'_i$ is
	renamed $p_i$ for all $i \in \int{n}$, and the ports accordingly. 
	Remark that $\oplus$ is associative but not commutative.
	
	We define a notation for gluing several graphs. 
	Let $\Gamma = \{G_i\}_{i \in I}$ be a finite family of $k$-graphs. For 
	$\omega = \omega_1\dots\omega_{\mu}$ a nonempty word on alphabet $I$, 
	the operation $\Delta^\Gamma(\omega)$ is defined by induction on the length 
        $\mu \in \N_+$ of $\omega$ as follows:
	\[
	  \Delta^{\Gamma}(\omega_1) = G_{\omega_1}
          \quad\text{and}\quad 
	  \Delta^{\Gamma}(\omega_1 \dots \omega_{\mu}) = 
	  \Delta^{\Gamma}(\omega_1 \dots \omega_{\mu-1}) \oplus G_{\omega_{\mu}}.
	\]
        In our main constructions, finite $k$-graphs will be glued to obtain a dynamics $\dyna{F}$.\linebreak

        \begin{remark}
       	  In the deterministic setting, we will always consider specific graphs such that,
          when glued together to produce the graph of a dynamics, we obtain a graph of out-degree one
          corresponding to a deterministic AN (Section~4 of \cite{glp24}).
        \end{remark}


	\paragraph{Logspace}
	Logspace is the complexity class of problems that
        can be solved with a deterministic Turing machine in space $\O(\log n)$ for inputs of size $n$.
        The space bound only applies to working tapes
        (the read-only input tape and write-only output tapes are not bounded,
        in particular to be able to read the input entirely)~\cite{perifel2014complexite}.
	For example, copying a circuit only requires one bit of memory,
        and manipulating object sizes (such as the input size) is possible in logspace.

	\section{Description of the metareduction}
        \label{s:metareduction}

	Rice-like complexity lower bounds for {\psiD} have been established in~\cite{ggpt21}
	(for deterministic SGRs and FO formulas)
	and in~\cite{ggpt23} (for non-deterministic SGRs and MSO formulas),
	and their results are strengthened in~\cite{glp24}
	from succinct graph representations to $q$-uniform automata networks for any $q\geq 2$.
	In the deterministic (resp.~non-deterministic) setting it states that for any $q$-non-trivial (resp.~$q$-arborescent)
	MSO formula $\psi$, the problem {\psiDq} (resp.~{\psiNDq}) is either $\NP$-hard or $\coNP$-hard.
	The $\NP$/$\coNP$ symmetry in this statement is necessary (unless $\NP=\coNP$),
	because some formulas are known to be $\NP$-complete
	(\emph{e.g.}~deciding the existence of a fixed point: $\exists x, x\to x$~\cite{a85})
	while others are known to be $\coNP$-complete
	(\emph{e.g.}~deciding whether it is the identity: $\forall x, x\to x$~\cite{p22}).

	All these hardness proofs consist in polynomial time many-one reductions from {\SAT},
	and the purpose of the present work is to explain them, meanwhile proving that they can be computed in logspace.
	Let $q\geq 2$ be a fixed alphabet size and $\psi$ be a fixed $q$-non-trivial (resp.~$q$-arborescent) MSO formula,
	hence defining a problem {\psiDq} (resp.~{\psiNDq}).
	We consider the stronger general construction from~\cite{glp24}
	which, given an instance $S$ of {\SAT}, assembles the dynamics $\dyna{F}$ of a $q$-uniform AN
	by gluing five fixed $k$-graphs denoted $\Gamma=\{G_0,G_1,G_2,G_3, G_4\}$.
	It is important to underline that the value of $k$ and the $k$-graphs of $\Gamma$ depend only on the formula $\psi$
	defining the problem, therefore they are of constant size for the purpose of the metareduction.
        We also have $|G_0|=|G_1|$.
	They are obtained using tools from finite model theory, and we refer to these articles for their definition
	(here their existence suffices).

	For $S$ a propositional formula with $s$ variables, we denote $\overline{S}$ the word of length $2^s$ such that 
	$\overline{S}_i$ is $1$ if $S(i)$ is false, and $0$ if $S(i)$ is true (viewing the binary expansion 
	of $i$ as a Boolean assignment for $S$).
	Then the metareduction consists in building the circuit encoding of a $q$-uniform AN (resp.~NAN)
        whose dynamics is (see~\cite[Section~6.3]{glp24}):
        \begin{align}\label{eq:dynaF}
	  \dyna{F} = \Delta^{\Gamma}(2 \cdot \overline{S} \cdot 4^{L(s)}\cdot 3)
        \end{align}

        \noindent
	where $L(s)=\frac{b}{a}(q^{(s+\log_q(\alpha)+1)\mu}-1)-\alpha 2^s\geq 0$
	and $a,b,\mu,\alpha,\log_q(\alpha)$ are constant integers depending only on $\psi$.
	This $\dyna{F}$ is a model of $\psi$ if and only if $G_0$ appears at least once, hence if and only if $S$ is satisfiable.
	The purpose of adding $L(s)$ copies of $G_4$ is to obtain a dynamics
	on $|\dyna{F}|=q^N$ vertices (configurations) for some $N\in\N$,
	hence consistent with the $q$-uniformity requirement.
        The main achievement of~\cite{glp24} is indeed to resolve these arithmetical constraints.
        From Equation~\ref{eq:dynaF}, summing graph sizes and subtracting merged port vertices, we obtain:
        \begin{align}\label{eq:N}
	  |\dyna{F}|=q^N=|G_2|+2^s\,|G_0|+L(s)\,|G_4|+|G_3|-k\,(2^s+L(s)+1).
        \end{align}
	As we will deal with these quantities in the logspace reduction,
	it is important to note the following (although it is straightforward from \cite{glp24}).

        \begin{proposition}\label{prop:LsN}
	  The binary values of $2^s$, $L(s)$, $N$ and $q^N$ are computable in space $\O(\log s)$.
	  Moreover, $\log_2(L(s))$ and $\log_2(q^N)$ are in $\O(s)$.
        \end{proposition}

	In the next section, we will explain the circuit of the deterministic
        (resp.~non-deterministic) automata network $F$
	constructed as an output of the reduction transformation, given $S$ as input.
	This requires meticulous care and precise notations defined in the rest of this section.
	During the construction of the circuit encoding $F$, we
	will allocate each configuration label from $0$ to $q^N-1$ to some copy
	of a graph of $\Gamma$ and explain how the circuit decodes a configuration
	$c \in \{0,\dots,q^N-1\}$ in order to identify which graph of $\Gamma$ it belongs to,
	and the relative position of that vertex in this graph
	(in particular, whether or not it is among the port vertices).
	With these two information, the circuit will be able to compute
	the image (resp.~set of images) $F(c)$.

	The construction of $\dyna{F}$ consists in gluing a total of $2^s+L(s)+2$ copies of the graphs $G_0, G_1, G_2, G_3$ and $G_4$.
	We denote $(H_i)_{i \in \{0,\dots,2^s+L(s)+1\}}$ the graphs such that
	$\dyna{F} = H_{0} \oplus H_1 \oplus \ldots \oplus H_{2^s +L(s)+1}$:
	\begin{itemize}[nosep]
	  \item $H_{0}=G_2$,
	  \item $H_i\in\{G_0,G_1\}$ depending on $i \in \{1,\dots,2^s\}$, and
	  \item $H_i=G_4$ for $i \in \{2^s+1,\dots,2^s+L(s)\}$,
	  \item $H_{2^s +L(s)+1}=G_3$.
	\end{itemize}
	
	For every $G_j$, its ports are written $P_1(G_j)$ and $P_2(G_j)$.
	From~\cite{glp24} the ports of $G_0$ and $G_1$ have the same labels as the ones of $G_4$
	(precisely, we have $G_1 = \Delta^{\{G_4\}}(4^{\alpha})$ and 
	$G_0 = G_4 \sqcup \tilde{G_0}$ where the ports are taken in $G_4$).
        For convenience, we consider that the ports $P_1(G_3)$
        have the same labels as the ports $P_1(G_4)$.

\begin{figure}[t]
          \centering\includegraphics{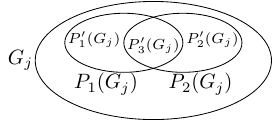}
	  \caption{
	    Notation for the ports of $G_j$ for $j\in\{0,1,4\}$.
	  }
	  \label{fig:ports}
	\end{figure}

	Recall that $P_1(G_j) \cap P_2(G_j)$ may be non-empty, and denote
	$P'_3(G_j) = P_1(G_j) \cap P_2(G_j)$,
	$P'_1(G_j) = P_1(G_j) \setminus P_3'(G_j)$ and
	$P'_2(G_j) = P_2(G_j) \setminus P_3'(G_j)$
	for $j \in \{0,1,4\}$.
	See Figure~\ref{fig:ports}.
	We define without ambiguity $P'_3 = P'_3(G_j)$ for any $j \in \{0,1,4\}$
	because in $\dyna{F}$ the vertices of $P'_3(G_j)$ for $j\in \{0,1,4\}$ will all be merged together.
	We have $P_1'(G_j) \cap P_2'(G_j) = \emptyset$ for $j \in \{0,1,4\}$.
	We extend the notation of $P'_1, P'_2, P'_3$ to $G_2$ and $G_3$
	as follows: $P'_3(G_2) \subseteq P_2(G_2)$ 
	and $P'_3(G_3) \subseteq P_1(G_3)$ such that these are the vertices that will be merge into $P'_3$
	in $\dyna{F}$. Also, set $P'_2(G_2) = P_2(G_2) \setminus P'_3(G_2)$ and 
	$P'_1(G_3) = P_1(G_3) \setminus P'_3(G_2)$. We define neither $P'_1(G_2)$ nor
	$P'_2(G_3)$ because the primary ports of $G_2$ and secondary ports of $G_3$ are not relevant.

	In the following we will use shortcuts for integer values relative to ports:
	\begin{align*}
		& g_j = |G_j| \textnormal{ for } j \in \{0,1,2,3,4\} \textnormal{ hence with } g_0=g_1\\
		&k_2 = |P_1'(G_j)| = |P_2'(G_j)| \textnormal{ for any } j \in \{0,1,4\}\\
		&k_3 = |P'_3| = k - k_2\\
		&g'_1 = g_1 - 2k_2 - k_3 = |G_1 \setminus (P_1(G_1) \cup P_2(G_1))|\\
		&g'_2 = g_2 - k_2 - k_3 = |G_2 \setminus P_2(G_2)|\\
		&g'_3 = g_3 - k_2 -k_3 = |G_3 \setminus P_1(G_3)|\\
		&g'_4 = g_4 - 2k_2 - k_3 = |G_4 \setminus (P_1(G_4) \cup P_2(G_4))|
	\end{align*}
	
\begin{figure}[t]
		\centering
		\includegraphics[width=\textwidth]{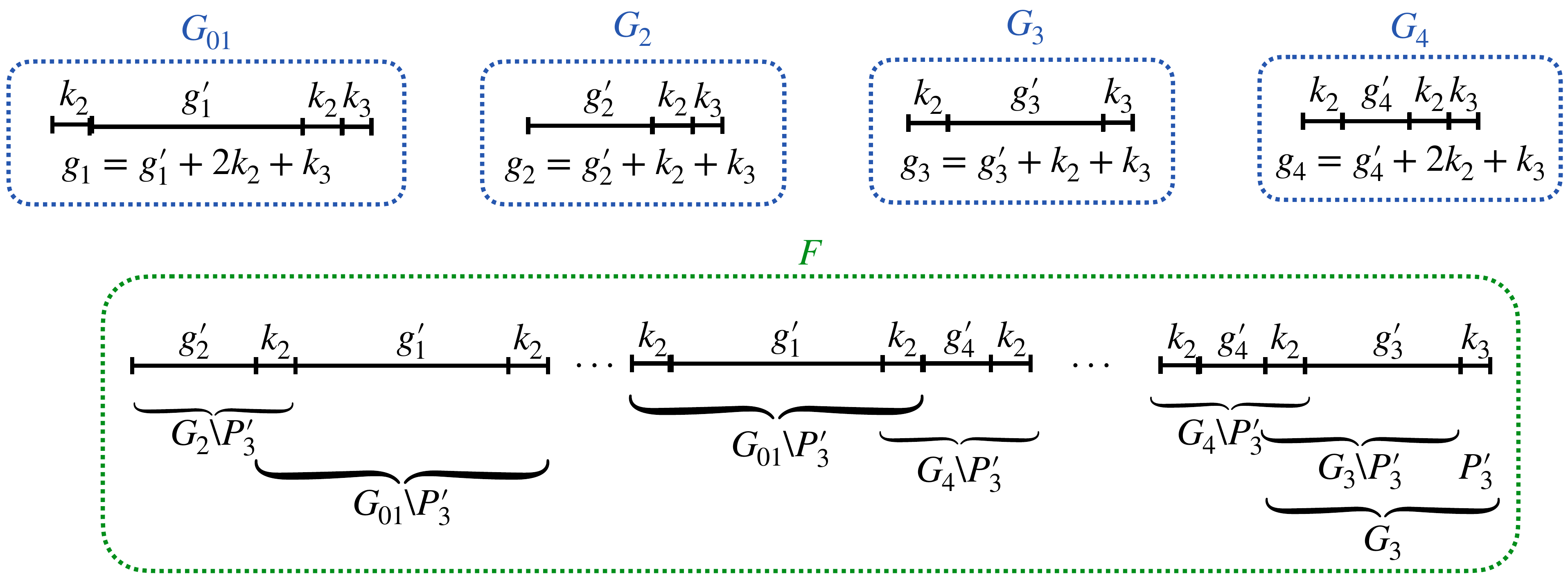}
		\caption{
			Allocation of the configurations of $F$ within the copies
			of $G_0, G_1, G_2, G_3, G_4$,
                        for the purpose of constructing the dynamics $\dyna{F}$ given by Equation~\ref{eq:dynaF}.
                        Recall that $G_{01}$ denotes $G_0$ or $G_1$ which have the same set of vertices (configurations).
                        Observe that $P'_3$, which are both primary and secondary ports that are merged in all copies,
                        appear at the very end of the configuration space of $F$.
		}
		\label{fig:configurations}
	\end{figure}

	Without loss of generality (because our MSO graph logic on signature $\{=, \to\}$
	is invariant by isomorphism),
	for any $j\in\{0,1,2,3,4\}$ we label the vertices of $G_j$ as $V(G_j)=\{0,\dots,g_j-1\}$, with
	$P'_1(G_j)=\{0,\dots,k_2-1\}$,
	$P'_2(G_j)=\{g_j-k_2-k_3,\dots,g_j-k_3-1\}$ and
	$P'_3=\{g_j-k_3,\dots,g_j-1\}$,
	except for $P'_1(G_2)$ and $P'_2(G_3)$ which are not necessary.
        We also denote $G_{01}$ for a graph which may be $G_0$ or $G_1$,
        depending on the evaluation of $S$ in Equation~\ref{eq:dynaF}
        (recall that $|G_0|=|G_1|$ and their configuration labels are compatible).
	See Figure~\ref{fig:configurations}.
	
	\begin{remark}
		In the following we will consider that we construct a $q$-uniform AN or NAN
                because by taking $G_4 = G_0$ and replacing $L(s)$ by any integer
		we still construct a circuit suitable for the reduction without the $q$-uniform 
		restriction.
	\end{remark}
	
	\section{Logspace circuit construction}
        \label{s:logspace}

        We now prove that for any MSO formula $\psi$ and any $q\geq 2$,
        the transformation from $S$ to $\dyna{F}$ described in Equation~\ref{eq:dynaF}
        of Section~\ref{s:metareduction} can be computed in logarithmic space.
        That is, the general $\NP$ or $\coNP$ hardness results on {\psiDq} and {\psiNDq}
        from~\cite{glp24} hold for logspace reductions.
        Recall that the five graphs of $\Gamma=\{G_0,G_1,G_2,G_3,G_4\}$ are constants depending only on the formula $\psi$
        defining the problem and that $L(s)$ and $N$ can be computed in logspace from $S$ (Proposition~\ref{prop:LsN}).
        Let $S$ be a propositional formula on $s$ variables.
        We first define the circuit of $F$ in the deterministic case, 
        and then adapt it in the non-deterministic case,
        while proving some useful properties about the circuit.
        Finally, we prove that these circuits can be computed in logspace.
        For each $\psi$, $q$ and case (deterministic or not) the graphs of $\Gamma$ are different,
        but they always respect the conventions established in Section~\ref{s:metareduction}.

        \subsection{Deterministic dynamics}
	
\begin{figure}[t]
		\centering
                \includegraphics[width=\textwidth]{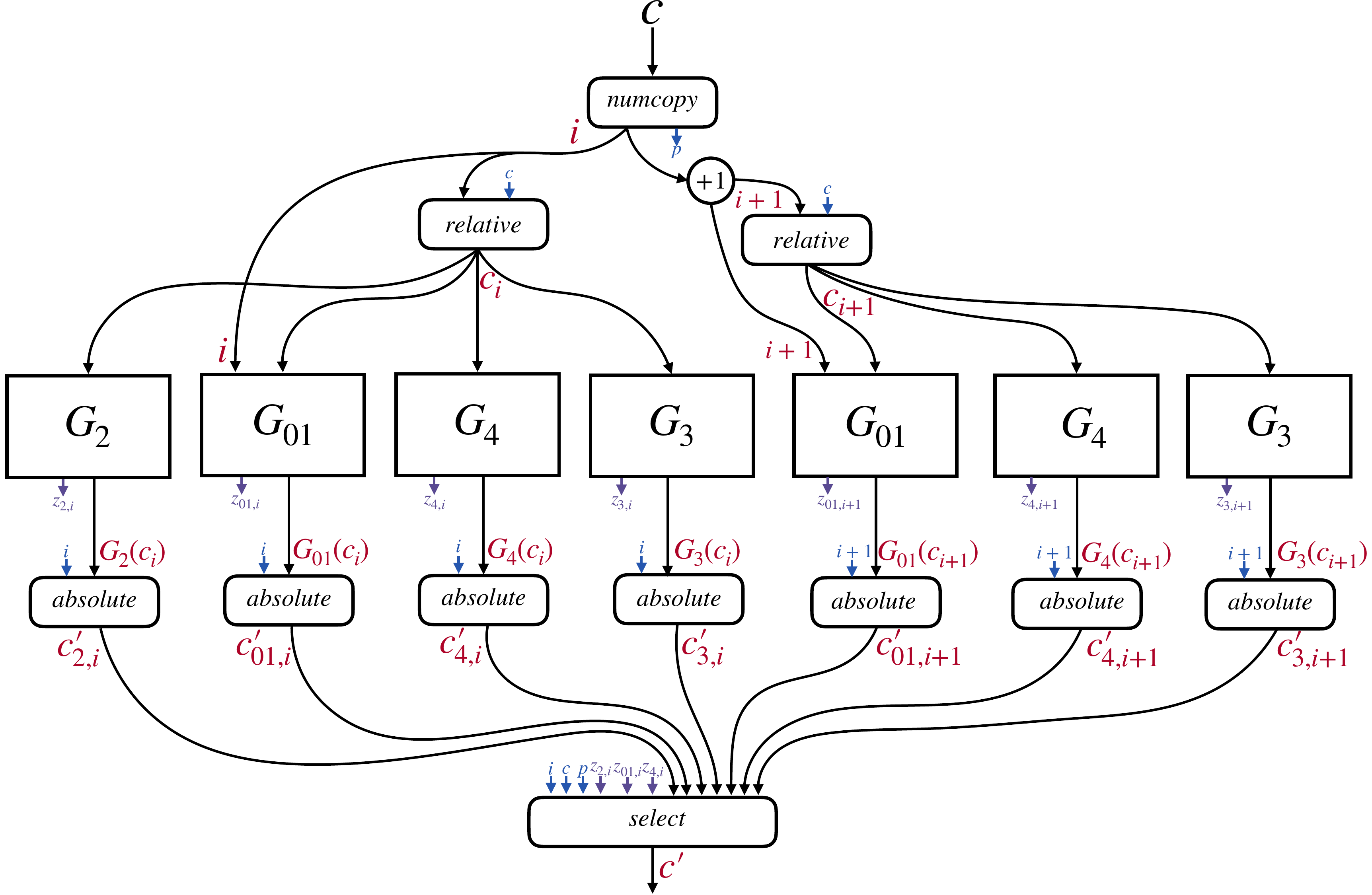}
		\caption{
                	Circuit of $F$, computing $F(c) = c'$. For sake of clarity, some wires are not 
			drawn, but are replaced with references in blue (for the wires coming from $c$
			and $i$ especially). We label in red the output of some subcircuits,
			these notations will be used in their formal description.
		}
		\label{fig:circuitDet}
	\end{figure}

        We now describe the circuit of the $q$-uniform AN $F$ with
        $N$ automata whose dynamics
        $\dyna{F}$ is described by Equation~\ref{eq:dynaF},
        \emph{i.e.} the output of the reduction from $S$ to prove that {\psiDq} is hard.
        In the deterministic case, each of the five graphs from $\Gamma$ have out-degree at most one
        and the result of their gluing also has out-degree exactly one (see~\cite[Section~4]{glp24},
        some port vertex may have no out-neighbor but it appears after gluing).
        Since they are constants (do not depend on $S$), computing their adjacencies takes $\O(1)$ space.
        Figure~\ref{fig:circuitDet} gives an overview of the construction,
	on $\lceil \log_2(q^N) \rceil$ input bits and $\lceil \log_2(q^N) \rceil$ output bits,
	with subcircuits represented as boxes. 
	Let $c$ be the input of the global circuit we are constructing.
	\begin{itemize}[nosep]
	  \item $numcopy$ identifies the index $i$ of the graph $H_i$ to which $c$ belongs.
	  \item $relative$ computes the label corresponding to $c$ in $H_i\in\Gamma$.
	  \item $G_j$ computes the out-neighbor of a given vertex label in $G_j$.
          \item $absolute$ is the inverse of $relative$.
	  \item $select$ selects among the subcircuits the correct out-neighbor of $c$.
	\end{itemize}
	The rest of this section provides a formal description of these subcircuits
	and an analysis of their complexity in order to eventually conclude
	that the whole circuit is computable in logspace from $S$.
	When the domain and range of a subcircuit are given, it means that other input words
	will have no effect on the final computation of the whole circuit
	and we do not even need to specify the output in this case.
	
	\subsubsection{Subcircuit $G_j$ for $j\in\{01,2,3,4\}$}
	
        Recall that the graphs from $\Gamma$ have out-degree at most one.
	For $j\in\{2,3,4\}$, the subcircuit $G_j:V(G_j)\to V(G_j)$ simply computes $G_j(c)$ on input $c$.
        It has an extra output bit $z$ equal to $1$ when $c$ has out-degree $0$ in $G_j$
        (that is, when $G_j(c)$ is undefined).
        When $z=1$, the other output bits supposed to encode $G_j(c)$ are set to $0$.
	Subcircuit $G_{01}:\{1,\dots,2^s\}\times V(G_0)\to V(G_0)$ has the same logic
        but produces a graph that is a copy of either $G_0$ or $G_1$
        (recall that $V(G_0)=V(G_1)$), depending on an evaluation of $S$.
        It has two inputs, $i$ and $c$:
	$$G_{01}(i, c)= 
	\begin{cases}
		G_1(c) &\textnormal{if } S(i-1) \textnormal{ is true},\\
		G_0(c) &\textnormal{otherwise.}
	\end{cases}$$
	For $i\in\{1,\dots,2^s\}$, it means that $H_i$ is a copy of $G_1$ if $S(i-1)$
        is true and a copy of $G_0$ if $S(i-1)$ is false.
        The binary value of $i$ is an input of the subcircuit $G_{01}$ and its output
        will only be selected (by subcircuit $select$) on domain $i\in\{1,\dots,2^s\}$,
        where the binary value of $i-1$ is a Boolean assignment for $S$.

        Given that the graphs have constant size and that the formula $S$ merely needs to be evaluated
        in order to branch on the case disjunction, it simply needs to be copied once inside
        the subcircuit $G_{01}$. Formally, the $j^\textnormal{th}$ output bit of $G_{01}(i,c)$
        is obtained as the following combination,
        using a copy of the circuit for $G_1$ evaluated on $c$ with $b_1$ its $j^\textnormal{th}$
        output bit, a copy of the circuit for $G_0$ evaluated on $c$ with $b_2$ its $j^\textnormal{th}$
        output bit, and a copy of $S$ evaluated on assignment $i-1$ (this bit is wired twice):
        \[
          (S(i-1) \wedge b_1) \vee (\neg S(i-1) \wedge b_2).
        \]
        This subcircuit can be assembled in logarithmic space
        as it only requires to copy $S$,
	and remember $|S|$ to place the decrement and constant circuits for $G_0$ and $G_1$ around,
        then add gates and wires for the final combination above.
		
	\subsubsection{Subcircuit $numcopy$}

	The subcircuit $numcopy: \{0, \dots, q^N-1\} \to \{0, \dots, 2^s +L(s)+1\} \times \{0,1\}$ computes, for each
	configuration $c$ of $\dyna{F}$, the smallest integer $i$ such that $c$ is in $H_i$,
	which is the $(i+1)^{\textnormal{th}}$ copied graph in the definition of $\dyna{F}$
	in Equation~\ref{eq:dynaF}
	(remark that it does not mean it is the only copied graph of $\dyna{F}$ that contains $c$
	because gluing operations merge vertices).
        It has an extra output bit $p$ equal to $1$ when $c$ is a secondary port in $H_i$.
        According to the nomenclature of Figure~\ref{fig:configurations}, we have $numcopy(c)=(i,p)$ with
        $i$ equal to:
	$$\begin{cases}
		0 &\textnormal{if } c < g'_2 + k_2 \textnormal{ or } c \ge q^N-k_3, \\
		\Big\lfloor \frac{c-g'_2-k_2}{g'_1 +k_2}\Big\rfloor &\textnormal{if } g'_2 + k_2 \le c < g'_2 + k_2 + 2^s(g'_1 + k_2), \\[.4em]
		\Big\lfloor \frac{c-g'_2-k_2 -2^s(g'_1+k_2)}{g'_4 +k_2}\Big\rfloor  + 2^s 
		&\textnormal{if } g'_2 + k_2 + 2^s(g'_1 + k_2) \le c < q^N-(k_2+g'_3+k_3), \\ 
		2^s +L(s) + 1 &\textnormal{if } q^N-(k_2+g'_3+k_3) \le c < q^N-k_3,
	\end{cases}$$
        and $p=0$ if and only if $c$ is part of an interval whose size is labeled by
        $g'_2$ or $g'_1$ or $g'_4$ or $g'_3$ on Figure~\ref{fig:configurations}
        (implemented as a straightforward case disjunction).

	The subcircuit for $numcopy$ can be constructed in logspace because the case disjunction
	relies on comparisons between the input $c$ and integer values of linear size (this is important to underline
	as we are required to construct in logspace the circuits performing the comparisons)
	and computable in logspace by Proposition~\ref{prop:LsN}.
	Moreover, a circuit for the division
        (which is also computed by the circuit itself, because it depends on $c$,
        even though the denominator is a constant).
	of length-$\O(s)$ binary numbers ($s$ is given in unary with $S$)
        can be constructed in logspace (see for example~\cite{w87}, and~\cite{cdl01} for a recently
        improved logspace-uniform family of circuits for the division).
	Combining these elements as a unique circuit again requires to manipulate their number of gates
	in order to compute the adjacency matrix of disjoint unions
	(in terms of graph) and then wire them (\emph{i.e.}, add arcs),
	but all these quantities are polynomial in $s$ therefore encoded on $\O(\log(s))$ bits.

	\subsubsection{Subcircuit $relative$}

	The subcircuit
	$relative: \{0, \dots, 2^s +L(s)+1\} \times \{0, \dots, 2^N-1\} \to V(G_j)$
	computes, for each number of copy $i$ and configuration $c$ in $\dyna{F}$,
	the label of $c$ within the graph $H_i$, hence with range in the vertices of $G_j$
	such that $H_i=G_j$.
	So $relative(i, c)$ equals:
	$$\begin{cases}
		c &\textnormal{if } i=0 \textnormal{ and } c < g'_2+k_2, \\
		c-g'_2-k_2-2^s(g'_1+k_2)-L(s)(g'_4+k_2)-g'_3 &\textnormal{if } i=0 \textnormal{ and } c \geq q^N - k_3, \\
		c - g'_2 - k_2 - (i-1)(g'_1 + k_2) &\textnormal{if } 1\leq i\leq 2^s, \\
		c - g'_2 - k_2 - 2^s(g'_1 + k_2) - (i-2^s-1)(g'_4+k_2) &\textnormal{if } 2^s<i\leq 2^s+L(s), \\
		c - g'_2 - k_2 - 2^s(g'_1 + k_2) - L(s)(g'_4 + k_2) &\textnormal{if } i=2^s+L(s)+1.
	\end{cases}$$
	The two first cases correspond to $i=0$, \emph{i.e.}~$H_i=G_2$,
	and they branch on the fact that $c$ belongs to $P'_3$ or not,
	in accordance with our choice of placing $P'_3$ at the end of the configuration labels in $\dyna{F}$.
	Observe that $relative$ will work as expected, because in Figure~\ref{fig:circuitDet}
        it is the output $i$ of $numcopy(c)$, or $i+1$, that is given as an input to $relative$,
        hence this subcircuit will output $relative(numcopy(c),c)$, or $relative(numcopy(c)+1,c)$.
        In particular, the case where $i=0$ and $g'_2+k_2\leq c<q^N-k_3$ never occurs
        as an input to $relative$, therefore we do not need to define this case.
        We will use $relative$ on $i$ and $i+1$ to handle the fact that,
        when gluing two graphs $H_i$ and $H_{i+1}$, a secondary port $v$ from $H_i$ may have, in $\dyna{F}$,
        its out-neighbor defined in $H_{i+1}$.
        Indeed, when $p=1$, that is when $c$ is a secondary port vertex in $H_i$,
        the subcircuit $relative(numcopy(c)+1,c)$ outputs the primary port vertex of $H_{i+1}$
        that will be merged with $c$.


	
	By Proposition~\ref{prop:LsN}, the subcircuit $relative$ is constructible in logspace.

	\subsubsection{Subcircuit $absolute$}

        The subcircuit $absolute: \{0,\dots, 2^s +L(s)+1\} \times V(G_j) \mapsto \{0,\dots,q^N-1\}$
        computes, for each number of copy $i$ and configuration $c$ within $H_i=G_j$,
	the label of $c$ in $\dyna{F}$.
	The output of $absolute(i,c)$ is defined as:
	$$\begin{cases}
		c &\textnormal{if } i=0 \textnormal{ and } c < g'_2 + k_2, \\
		c + g'_2 + k_2 + 2^s(g'_1 + k_2) + L(s)(g'_4 + k_2) + g'_3 &\textnormal{if } i=0 \textnormal{ and } c \geq g'_2 + k_2, \\
		c + g'_2 + k_2 + (i-1)(g'_1 + k_2) &\textnormal{if } 1\leq i\leq 2^s, \\
		c + g'_2 + k_2 + 2^s(g'_1 + k_2) + (i - 2^s - 1)(g'_4 + k_2) &\textnormal{if } 2^s<i\leq 2^s + L(s), \\
		c + g'_2 + k_2 + 2^s(g'_1 + k_2) + L(s)(g'_4 + k_2) &\textnormal{if } i=2^s+L(s)+1.
	\end{cases}$$
	Remark that for any given $i$ the subcircuit $absolute$ computes the inverse of
	$relative$, in the sense that we have $absolute(i,relative(i,c))=c$.
	This is straightforward to check from their definition, except for $i=0$ and $g'_2+k_2\leq c<q^N-k_3$
	because $relative$ is not defined, but as argued above it never occurs.

	By Proposition~\ref{prop:LsN}, the subcircuit $absolute$ is constructible in logspace.

	
	

	\subsubsection{Subcircuit $select$ and assembly}
	
	Finally we present the function that will sort all the information computed
        in the assembly of previous subcircuits
        in order to output the correct out-neighbor of $c$ in $\dyna{F}$.
        Before giving a formal description of $select$ and argue that the whole circuit indeed computes $F$
        with $\dyna{F}$ given by Equation~\ref{eq:dynaF},
        let us underline that we will consider the wirings given on Figure~\ref{fig:circuitDet},
        and the output labels it defines.
        The subcircuit:
        $$select:\{0, \dots, 2^s +L(s)+1\} \times \{0, \dots, q^N-1\}^8
        \times \{0,1\}^4 \to  \{0, \dots, q^N -1\}$$
        takes as inputs:
        \begin{itemize}[nosep]
          \item $i$ and $c$ such that $c$ is in $H_i$ (\emph{i.e.}, $i$ is the output of $numpcopy(c)$),
          \item the outputs from all the subcircuits for the graphs
            (after conversion to $\dyna{F}$ through $absolute$),
            denoted $c'_{2,i}, c'_{01,i}, c'_{4,i}, c'_{3,i}, c'_{01,i+1}, c'_{4,i+1}, c'_{3,i+1}$,
          \item the bits $p$ (whether $c$ is a secondary port in $H_i$) and
            $z_{2,i},z_{01,i},z_{4,i}$
            (whether $c$ has out-degree zero as a primary port in $G_2,G_{01},G_4$, respectively).
        \end{itemize}
        The purpose of $select$ is to output only one configuration
        among the seven possible configurations
        ($c'_{2,i}, c'_{01,i}, c'_{4,i}, c'_{3,i}, c'_{01,i+1}, c'_{4,i+1}, c'_{3,i+1}$):
        the out-neighbor of $c$ in $\dyna{F}$.

        Formally, the subcircuit $select$ outputs:
	\[
	  c'=
	  \begin{cases}
	    c'_{2,i} &\textnormal{if } i=0 \textnormal{ and } c<g'_2+k_2 \textnormal{ and } (p=0 \textnormal{ or } z_{2,i}=0), \\
	    c'_{01,i+1} &\textnormal{if } i=0 \textnormal{ and } c<g'_2+k_2 \textnormal{ and } p=1 \textnormal{ and } z_{2,i}=1, \\
	    c'_{01,i} &\textnormal{if } 1\leq i\leq 2^s \textnormal{ and } (p=0 \textnormal{ or } z_{01,i}=0), \\
	    c'_{01,i+1} &\textnormal{if } 1\leq i<2^s \textnormal{ and } p=1 \textnormal{ and } z_{01,i}=1, \\
	    c'_{4,i+1} &\textnormal{if } i=2^s \textnormal{ and } p=1 \textnormal{ and } z_{01,i}=1, \\
	    c'_{4,i} &\textnormal{if } 2^s<i\leq 2^s+L(s) \textnormal{ and } (p=0 \textnormal{ or } z_{4,i}=0), \\
	    c'_{4,i+1} &\textnormal{if } 2^s<i<2^s+L(s) \textnormal{ and } p=1 \textnormal{ and } z_{4,i}=1, \\
	    c'_{3,i+1} &\textnormal{if } i=2^s+L(s) \textnormal{ and } p=1 \textnormal{ and } z_{4,i}=1, \\
	    c'_{3,i} &\textnormal{if } i=2^s+L(s)+1,\\
	    c'_{2,i} &\textnormal{if } i=0 \textnormal{ and } c\geq q^N-k_3 \textnormal{ and } z_{2,i}=0, \\
	    c'_{4,i+1} &\textnormal{if } i=0 \textnormal{ and } c\geq q^N-k_3 \textnormal{ and } z_{2,i}=1 \textnormal{ and } z_{4,i+1}=0, \\
	    c'_{3,i+1} &\textnormal{if } i=0 \textnormal{ and } c\geq q^N-k_3 \textnormal{ and } z_{2,i}=1 \textnormal{ and } z_{4,i+1}=1. \\
	  \end{cases}
	  \quad
	  \tikz[baseline=-9.5em]{
	    \def\hauteurducase{0.505}
	    \foreach \y in {1,...,12}
	      \node at (0,-\y*\hauteurducase) {($\y$)};
	  }
	\]
	In this case disjunction, the cases on lines 1--2 handle $H_0$ excepted $P'_3$:
	if $c$ is a secondary port ($p=1$) and $c$ has no out-neighbor in $H_0$ (which is a copy of $G_2$),
	then its out-neighbor is taken from $H_1$ (which is a copy of $G_{01}$).
	The cases on lines 3--5 handle $H_i$ for $i\in\{1,\dots,2^s\}$:
	if $c$ is a secondary port ($p=1$) and $c$ has out-neighbor in $H_i$ (which is a copy of $G_{01}$),
	then its out-neighbor is taken from $H_{i+1}$ (which is a copy of $G_{01}$ when $i\neq 2^s$, and a copy of $G_4$ otherwise).
	The cases on lines 6--8 handle $H_i$ for $i\in\{2^s+1,\dots,2^s+L(s)\}$:
	if $c$ is a secondary port ($p=1$) and $c$ has out-neighbor in $H_i$ (which is a copy of $G_4$),
	then its out-neighbor is taken from $H_{i+1}$ (which is a copy of $G_4$ when $i\neq 2^s+L(s)$, and a copy of $G_3$ otherwise).
	The case on line 9 handles $H_{2^s+L(s)+1}$:
	configuration $c$ is never a secondary port in this case (hence it is not necessary to precise that $p=0$),
	and the out-neighbor is always in $H_{2^s+L(S)+1}$ (which is a copy of $G_3$).
	Finally, the cases lines 10--12 handle $P'_3$:
	configuration $c$ is always a secondary port in this case (hence it is not necessary to precise that $p=1$),
	and if its out-neighbor is not in $H_0$ (which is a copy of $G_2$),
	then it may be either be within $P'_3$ from the primary ports of $H_1$ (computed from a copy of $G_4$,
	which has the same port vertices as $G_0$ and $G_1$),
	or the out-neighbor is in $H_{2^s+L(S)+1}$ (which is a copy of $G_3$, whose port labels are identical to $G_4$,
	so that it is convenient here to also use $c_{i+1}$ in the subcircuit for $G_3$).

	From the explanations above, we conclude that the whole circuit computes
	$F$ such that $\dyna{F}$ is defined in Equation~\ref{eq:dynaF}.
	The argumentation relies on the fact that this $\dyna{F}$ from~\cite{glp24} has, by hypothesis,
	out-degree exactly one, hence if a secondary port vertex has out-degree zero then its out-neighbor necessarily comes
	from the primary port vertex merged with it during the gluing operations
        (and if a secondary port vertex has an out-neighbor in two graphs $H_i,H_{i+1}$,
        then these two out-neighbors will be merged and our circuit considers the first one).
	The subcircuit for $select$ is again a case disjunction with simple arithmetics
	on $\O(s)$-bits integers, constructible in logspace.

	\begin{remark}
	  If $L(s)=0$ then there is no copy of $G_4$, and $select$ is simply patched by replacing $c'_{4,i+1}$ line 5 with $c'_{3,i+1}$,
	  and deleting line 8 to avoid an ambiguity for $i=2^s$ (lines 6--7 are vacuous, and line 11 is still correct).
	\end{remark}

	\subsubsection{Logspace reduction}
	
	Each subcircuit has a size polynomial in $s$, hence encoded on $\O(\log(s))$ bits,
	therefore there is no difficulty in building a unique adjacency matrix (and gates vector)
	with additional wires for their assembly (between subcircuits inputs and outputs).
	For example, the $\lceil\log_2(q^N)\rceil$ output bits of each $absolute$ can be grouped as consecutive gates,
	and remembered as a logspace pointer to the first of these output gate.
	There are finitely many such grouped wirings to implement.
	We conclude that, from an instance $S$ of {\SAT}, the circuit described on Figure~\ref{fig:circuitDet}
        can be constructed in logspace. Theorem~\ref{theorem:main} then follows from~\cite{glp24}
        (Section~\ref{s:metareduction}) regarding {\psiDq}.

	
	\section{Non-deterministic dynamics}
	\label{s:nondet}
	
\begin{figure}[t]
		\centering
		\includegraphics[width=\textwidth]{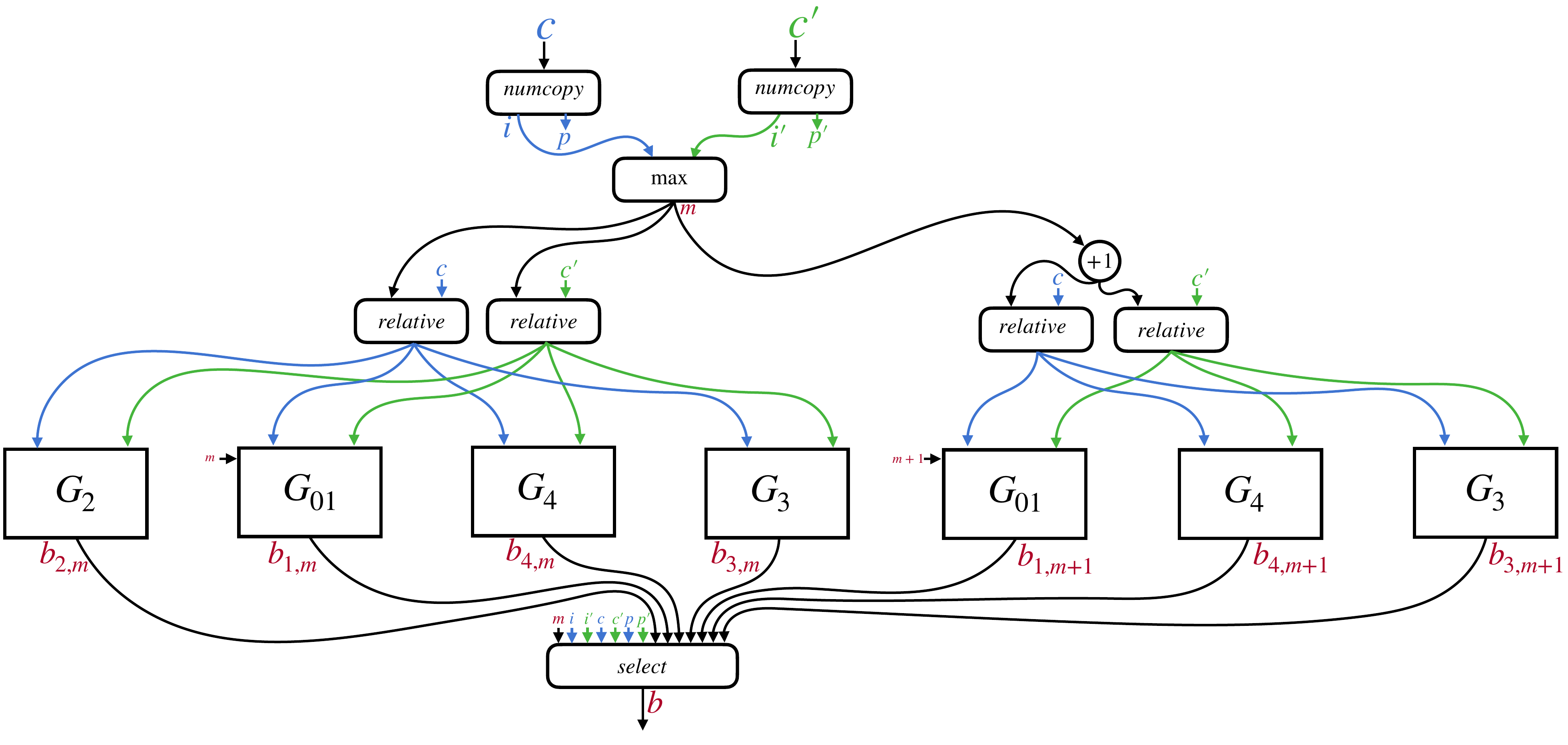}
		\caption{
			Circuit of $F$, computing $F(c, c') = b$. For clarity,
			some wires are not drawn, but are replaced with references in blue (relative to $c$),
			green (relative to $c'$) and red.
		}
		\label{fig:circuitNonDet}
	\end{figure}

	In this section we will show how to construct a circuit for the $q$-uniform AN $F$ with
	$N$ automata
	whose dynamics is given by Equation~\ref{eq:dynaF},
	in the non-deterministic case.
	The graphs from $\Gamma$ and $\dyna{F}$ are arbitrary.
	Recall that when port vertices are merged, all arcs are kept. 
	Figure~\ref{fig:circuitNonDet} gives an overview of the construction,
	on $2\,\lceil\log_2(q^N)\rceil$ input bits and $1$ output bit,
	with subcircuits again represented as boxes
	(the circuits of the graphs from $\Gamma$ also take a pair of configurations as input,
	and output the corresponding adjacency bit,
	\emph{i.e.}~whether there is an arc from the first to the second).
	
	We still use the exact same functions as in the deterministic case 
	for $numcopy$ and $relative$.
	We will adapt the definition of $G_{01}$, $select$, but most importantly we will
	introduce the function $max$ to handle the case where $c$ and $c'$ belong to different copies.
	
	\subsubsection{Subcircuit $G_{01}$}
	
	As in the deterministic case, $G_{01}$ is still a copy of either $G_0$ or $G_1$,
	depending on an evaluation of $S$. Now it has three inputs:
	$$G_{01}(i, c, c')= 
	\begin{cases}
		G_1(c, c') &\textnormal{if } S(i-1) \textnormal{ is true},\\
		G_0(c, c') &\textnormal{otherwise}.
	\end{cases}$$
	
	\subsubsection{Subcircuits $max$ and $select$}
	
	One difficulty that can arise in the non-deterministic case, is that given two 
	configurations $c$ and $c'$ as an input, when computing $numcopy(c)=(i,p)$
	and $numcopy(c')=(i',p')$ we have $i \neq i'$.
	Our goal in the circuit of $F$ 
	is to verify whether there exists a graph $H_j$ such that $H_j(c, c')=1$,
	and for this purpose it is convenient to introduce $m = max(i,i')$.
	We proceed to a case analysis centered on ports
	(recall $p,p'$ indicate whether $c,c'$ are secondary ports, respectively),
	and indicate the corresponding cases in the next definition of $select$
	which is centered on the value of $m$ (and sometimes $m+1$).
	
	\begin{itemize}[nosep]
		\item If neither $c$ nor $c'$ belongs to $P'_3$, then we need to search for the existence of the arc
		$(c, c')$ in at most two graphs.
		\begin{itemize}[nosep]
			\item If $i=i'=m$ and ($p=0$ or $p'=0$), then
			we search in $H_{m}$ (cases 3--7).
			\item If $i=i'=m$ and $p=1$ and $p'=1$, then
			we search in $H_{m}$ and $H_{m+1}$ (cases 3--7).
			\item If $i'=i+1=m$ and $p=1$, then we search in $H_{m}$ (cases 8--10).
			\item If $i=i'+1=m$ and $p'=1$, then we search in $H_{m}$ (cases 8--10).
			\item Otherwise $|i-i'|>1$, or $|i-i'|=1$ and the configuration in $H_{\min(i,i')}$
			is not a secondary port, so no graph contains both vertices (case 1).
		\end{itemize}
		\item If $c$ is in $P'_3$ and $c'$ is not in $P'_3$, then $i=0$, $m=i'$,
		and it depends on whether $c'$ is a secondary port (exists in two graphs) 
		(cases 3--7).
		\begin{itemize}[nosep]
			\item If $p'=0$ then we search only in $H_{m}$.
			\item If $p'=1$ then we search in $H_{m}$ and $H_{m+1}$.
		\end{itemize}
		\item If $c$ is not in $P'_3$ and $c'$ is in $P'_3$, then $i'=0$, $m=i$,
		and it depends on whether $c$ is a secondary port (exists in two graphs)
		(cases 3--7).
		\begin{itemize}[nosep]
			\item If $p=0$ then we search only in $H_{m}$.
			\item If $p=1$ then we search in $H_{m}$ and $H_{m+1}$.
		\end{itemize}
		\item If $c,c'$ are both in $P'_3$, then
		we search in all graphs (by construction, $c$ and $c'$ have the same labels in all graphs) (case 2).
	\end{itemize}
	
	
	The subcircuit:
	\[
	select:\{0,\dots,2^s+L(s)+1\}^3 \times \{0,\dots,q^N-1\}^2 \times \{0,1\}^9 \to \{0,1\}
	\]
	takes as inputs:
	\begin{itemize}[nosep]
		\item $i,i'$ and $c,c'$ such that the former are the outputs of $numcopy$ on the latter,
		\item the output bits from all the subcircuits for the graphs of $\Gamma$,
		denoted\linebreak $b_{2,m},b_{01,m},b_{4,m},b_{3,m},b_{01,m+1},b_{4,m+1},b_{3,m+1}$,
		\item the bits $p$ and $p'$ (whether $c$ and $c'$ are secondary ports of $H_i$ and $H_{i'}$, respectively).
	\end{itemize}
	It follows the case analysis presented above
	and factorizes $H_m$ and $H_{m+1}$ according to the value of $m$, with the search for
	an arc in multiple graphs implemented as a logical \emph{or} (denoted $\vee$).
	Recall that a vertex belongs to $P'_3$ when its label is greater or equal to $q^N-k_3$, hence $select$ outputs $b$ equal to:
	
	%
	%

	{\small
		\[
		\begin{cases}
			0 &\textnormal{if } |i-i'| > 1 \textnormal{ and } c < q^N-k_3 \textnormal{ and } c' < q^N-k_3,\\
			
			b_{2,m} \vee b_{01,m} \vee b_{4,m}
			&\textnormal{if } c \ge q^N-k_3 \textnormal{ and } c' \ge q^N-k_3,\\
			\quad
			\vee b_{01,m+1} \vee b_{4,m+1} \vee b_{3,m+1}\\
			
			b_{01,m} \vee (b_{01,m+1} \wedge p \wedge p')
			&\textnormal{if } 0 < i = i' < 2^s \textnormal{ or } (c \ge q^N-k_3 \textnormal{ and } 0 < i' < 2^s)\\
			&\phantom{\textnormal{if } 0 < i = i' < 2^s }
			\textnormal{ or } (c' \ge q^N-k_3 \textnormal{ and } 0 < i < 2^s),\\
			
			b_{4,m} \vee (b_{4,m+1} \wedge p \wedge p')
			&\textnormal{if } 2^s < i = i' < 2^s+L(s)\\
			&\textnormal{or } (c \ge q^N-k_3 \textnormal{ and } 2^s < i' < 2^s+L(s))\\
			&\textnormal{or } (c' \ge q^N-k_3 \textnormal{ and } 2^s < i < 2^s +L(s)),\\
			
			b_{01,m} \vee (b_{4,m+1} \wedge p \wedge p')
			&\textnormal{if } i = i' = 2^s \textnormal{ or } (c \ge q^N-k_3 \textnormal{ and } i' = 2^s)\\
			&\phantom{\textnormal{if } i = i' = 2^s }
			\textnormal{ or } (c' \ge q^N-k_3 \textnormal{ and } i = 2^s),\\
			
			b_{2,m} \vee (b_{01,m+1} \wedge p \wedge p'))
			&\textnormal{if } (i=i'=0 \textnormal{ and } c < q^N-k_3 \textnormal{ and } c' < q^N-k_3)\\
			&\textnormal{or } (c \ge q^N-k_3 \textnormal{ and } i' = 0)\\
			&\textnormal{or } (c' \ge q^N-k_3 \textnormal{ and } i = 0),\\
			
			b_{4,m} \vee (b_{3,m+1} \wedge p \wedge p')
			&\textnormal{if } (i = i' = 2^s+L(s) \textnormal{ and } c < q^N-k_3 \textnormal{ and } c' < q^N-k_3)\\
			&\textnormal{or } (c \ge q^N-k_3 \textnormal{ and } i' = 2^s+L(s))\\
			&\textnormal{or } (c' \ge q^N-k_3 \textnormal{ and } i = 2^s+L(s)),\\
			
			b_{01,m} &\textnormal{if } |i-i'| = 1 \textnormal{ and } 0 < m \le 2^s,\\
			b_{4,m} &\textnormal{if } |i-i'| = 1 \textnormal{ and } 2^s < m \le 2^s+L(s),\\
			b_{3,m} &\textnormal{if } i = i' = 2^s+L(s)+1.
		\end{cases}
		\quad
		\tikz[baseline=-14.5em]{
			\def\hauteurducase{0.465}
			\newcounter{hextrac}
			\foreach \y/\hextra in {1/0,2/1,3/1,4/2,5/1,6/2,7/2,8/0,9/0,10/0}{
				\node at (0,-\y*\hauteurducase-\value{hextrac}*\hauteurducase) {($\y$)};
				\addtocounter{hextrac}{\hextra}
			}
		}
		\]
	}
	
	The final assembly and logspace reduction are as the deterministic case.
	

	\section{Conclusion and perspectives}
        \label{s:conclusion}
	 
	While detailing the circuit metaconstruction of Rice-like complexity lower bounds
	introduced in~\cite{ggpt23} for succinct graph representations
	and refined in~\cite{glp24} for $q$-uniform automata networks,
	we have demonstrated that it is feasible in logspace.
	The main non-constant part of the metaconstruction given
	by Equation~\ref{eq:dynaF} consists in evaluating $S$.
        Proving that this metareduction is computable in logspace
	may basically amount to exploiting the fact that {\CVP}
        (Circuit Value Problem) is $\Poly$-complete
        for many-one logspace reductions,
        but we have presented a much more direct construction of the circuit.
	Indeed, to formally exploit the fact that our circuits for $\dyna{F}$ implement polytime algorithms,
	we would have to explain how to construct in logspace (from $S$)
	an algorithm for the circuit outputted by the metareduction
	(on input $c$ in the deterministic case, or on input $c,c'$ in the non-deterministic case),
	which requires an almost equal care, but is less explicit.
	
	The final result of \cite{glp24}, whose reduction is improved in the current article, 
	is that {\psiDq} and {\psiNDq} are $\NP$-hard or $\coNP$-hard if 
	$\psi$ is non-trivial in the former case, or arborescent in the latter case.
	In the deterministic case, if $\psi$ is trivial, then
	 {\psiDq} can be solved in constant time. However, in the non-deterministic case,
	 if $\psi$ is not arborescent, then 
	 the complexity of {\psiNDq} is still open. 
	 The difference comes form the fact that an arborescent formula can be seen
	 as a non-trivial formula that has also infinitely many models and counter-models
	 of bounded treewidth. Nevertheless, dynamics of unbounded treewidth
	 may be pertinent, and the associated decision problem may be hard as well.
	 A track of research consists in considering other graph parameters,
	 such as cliquewidth or twinwidth, in order to push these results forward.

	As a consequence of our results, the problems
	{\psiDq} and {\psiNDq} are $\Poly$-hard (if $\psi$ is respectively non-trivial
	or arborescent).
	More interestingly, it showcases how to obtain $\Poly$-hardness
	with the technique of Rice-like metareduction for the study of the complexity
	of finite dynamical systems introduced in~\cite{ggpt23,glp24}.
	This is among current tracks of research for the development of Rice-like metatheorems,
	in particular for the introduction of new relations in the signature of monadic second order formulas.
	Indeed, introducing relations distinguishing configurations,
	for example a total or partial order $<$ on binary strings,
	results in breaking the isomorphism invariant of $\{=,\to\}$,
	that is the fact that $G\sim G'$ implies $G\models\psi \iff G'\models\psi$.
	It follows that ``easy'' questions can be asked,
	such as whether the minimum configuration $0^n$ is a fixed point:
	$\forall c: (\forall c': c\neq c' \implies c<c') \implies (c\to c)$.
	This question is obviously $\Poly$-complete, and we conjecture that a Rice-like result of the form
	``any non-trivial question on the signature $\{=,\to,<\}$ is $\Poly$-hard'' holds.

        Finally, providing an explicit construction of the circuits succinctly encoding
        the ANs produced by metareductions is also a ground step
        to the transfer of Rice-like complexity lower bound metatheorems
        to other models of computation
        (\emph{e.g.}~reaction systems, membrane models, DNA folding, finite cellular automata...),
        which may require to accommodate additional constraints.

	
        \paragraph{Acknowledgements}

        The authors thank projects
        ANR-24-CE48-7504 ALARICE,
        HO\-RIZON-MSCA-2022-SE-01 101131549 ACANCOS,
        STIC AmSud CAMA 22-STIC-02. 

	\bibliographystyle{splncs04} 
	\bibliography{biblio}

      \end{document}